# Probing Lattice Vibration at Surface and Interface of SiO$_2$/Si with Nanometer Resolution

Yuehui Li[1,2], Mei Wu[1,2], Ruishi Qi[1], Ning Li[1,2], Yuanwei Sun[1,2], Chenglong Shi[3], Xuetao Zhu[4], Jiandong Guo[4], Dapeng Yu[2,5,6], Peng Gao[1,2,6*]

[1]International Center for Quantum Materials, School of Physics, Peking University, Beijing, 100871, China

[2]Electron Microscopy Laboratory, School of Physics, Peking University, Beijing, 100871, China

[3]Nion Company, 11511 NE 118Th Street, Kirkland, Washington State, 98034, USA

[4]Beijing National Laboratory for Condensed Matter Physics and Institute of Physics, Chinese Academy of Sciences, Beijing 100190, China

[5]Shenzhen Key Laboratory of Quantum Science and Engineering, Shenzhen 518055, China

[6]Collaborative Innovation Centre of Quantum Matter, Beijing 100871, China.

Y.H.L. and M.W. contributed equally to this work.

E-mails: p-gao@pku.edu.cn

**Abstract** Recent advances in monochromatic aberration corrected electron microscopy make it possible to detect the lattice vibration with both high-energy resolution and high spatial resolution. Here, we use sub-10 meV electron energy loss spectroscopy to investigate the local vibrational properties at surface and interface of an amorphous $SiO_2$ (a-$SiO_2$) thin film on Si substrate. We find that each optical mode splits into three sub-modes, i.e., surface mode, bulk mode and interface mode, which can be measured from different locations. The pure surface modes can be measured in the vacuum near the surface, and the pure interface modes are expected to be obtained either at the interface location or in the Si, while inside the $SiO_2$ the measured signal is a mixture of bulk, surface, and interface modes. The bulk mode has the largest vibration energy and surface mode has the lowest. The energy of surface mode is thickness dependent, showing a blue-shift as z-thickness (parallel to fast electron beam) of $SiO_2$ film increases, while the bulk and interface modes have constant vibration energy. The intensity of bulk mode linearly increases with thickness being increased, and it drops steeply to zero near the surface and interface (within a few nanometers). The surface modes decay slowly in the vacuum following a Bessel function. The mechanism of the observed spatially dependent vibration behavior is discussed and quantitatively compared with dielectric response theory analysis. Our nanometer scale measurements of vibrations properties provide useful information about the bonding conditions at the surface and interface and thus may help to design better silicon-based electronic devices via surface and interface treatments.

**Introduction**

Lattice vibrations play a major role in many of the physical properties of condensed matter, including thermal (e.g. specific heat, thermal conductivity, and phase transition), electrical (e.g. electrical conductivity, superconductivity), and optical (e.g. polaritons, photonic, negative refraction) properties.[1] Measurements of vibration properties are vital to analyze the chemical bonding, isotope information, defect density, and strain state. At the surface and interface of solid materials, the change of atomic bonding state is expected to alter the local vibration behavior (such as phonon softening[2]), which should influence the localized physical properties via various phonon interactions (e.g. phonon-photon, phonon-electron, phonon-phonon). Therefore, exploring the localized vibration behavior is of significant importance. Particularly for the low dimensional materials in which the effects of surface and interface become more pronounced, the localized vibration properties may dominate the entire response of devices.[3,4]

The vibrational properties are commonly measured by high-resolution electron energy loss spectroscopy (HREELS),[5,6] inelastic neutron scattering, infrared spectroscopy (IR),[7,8] and Raman spectroscopy.[9] The spatial resolution of these techniques, however, is typically at micrometer-scale (unless assisted by a sharp metal tip, e.g. tip enhanced Raman spectroscopy,[10] ~20nm), and thus it is not convenient to extract the local vibration properties. On the other hand, the tip-enhanced techniques based on surface probe are inaccessible to the vibrational properties of the buried interface. In contrast, the recent advancements of dedicated scanning transmission electron microscope with aberration corrector and monochromatic cold field emission gun enable an electron probe with ~10 meV in energy resolution and ~0.1 nm in spatial resolution, allowing us to directly map the phonon space distribution of a single nanostructure,[11–16] to obtain the phonon structure,[17,18] to detect the temperature in the nanoenvironment,[19,20] and to prevent the radiation damage,[21] et.al. which provides an unprecedented opportunity to study the localized vibration properties of surface and interface.

In this paper, we use ~8 meV electron probe based on Nion UltraSTEM$^{TM}$ 200 microscope to study the vibrational properties of $SiO_2$/Si heterostructure. $SiO_2$ film is commonly used as the gate dielectric material in metal-oxide-semiconductor field effect transistors (MOSFETs). The quality and properties of the $SiO_2$ films and the $SiO_2$/Si interface can significantly influence the performance and stability of MOSFET's and bipolar transistors used in the integrated circuits.[22,23] The nanoscale structure and properties of the $SiO_2$/Si interface becomes more and more important as the device size drops to tens of nanometer or lower.[24] For example, previously IR spectroscopy showed that purely geometrical effects result in substantial red-shifts of the transverse-optical (TO) mode while no obvious thickness dependence for the longitudinal (LO) mode for slab thickness ranging from 1000 nm to 10 nm.[25,26] The recent electron microscopy studies[11,21] provided more valuable insights into the decay distance of some aloof modes and intensity change at the boundary. However, the limited energy resolution in the previous studies (typically ~16 meV in ref [11,21]) was not sufficient to distinguish the surface and other modes and thus the mode-dependent vibration behavior at the surface and interface is still largely unknown.

In our study, we improve the energy resolution up to ~8 meV at 60 kV, which allows us not only simultaneously acquire three different vibration modes [i.e., $TO_1$ (~60 meV), $TO_2$ (~103 meV), $TO_3$ (~ 130 meV–150 meV)], but also capture the subtle changes of vibration energy and intensity at the surface and interface of $SiO_2$ film on Si substrate. We find that the vibration is highly spatially dependent and each optical mode splits into three branches (surface modes, bulk modes, interface modes) that can be measured from different locations. The surface modes have the lowest energy and the bulk modes have the largest. Particularly, as z-thickness decreases (z denotes the electron propagation direction), the vibration energy of surface modes show a substantial red-shift that is attributed to the enhanced coupling of upper and lower surfaces,[27] while the energy of bulk and interface modes remain unchanged. The intensity of surface modes decay slowly in the vacuum following a Bessel function, while the intensity of bulk that is proportional to the slab thickness, drops steeply to zero near the surface and interface (within a few nanometers). Approaching the interface, the interface and guided modes cannot be distinguished due to the similar vibration energy, but their total intensity sharply decreases. These results can be qualitatively interpreted by a semiclassical relativistic local dielectric model.[27–29] Our study reveals the vibration behavior of $SiO_2$ at the surface and interface with nanometer resolution, providing useful insights into understanding the fundamental electrical, optical and thermal properties of surface and interface of $SiO_2$.

**Results and discussion**

A ~6 meV zero-loss peak can be tuned up at 60 kV (Supporting Fig S1). For a higher ratio of signal to noise, we significantly increase the acquisition time in practical measurements which only sacrifices the energy resolution slightly. Fig 1a shows three typical regions to record the spectra, corresponding to surface mode (blue), bulk plus surface mode (orange) and interface mode (cyan). Fig 1b shows the typical measured EELS spectra with a probe beam (with energy resolution ~8 meV) located near surface in vacuum side (cyan) and vacuum far away from specimen (green) where no vibration signal is detected. After background subtraction, the spectra in Fig 1c shows three peaks which are in agreement with previous IR spectroscopy and neutron scattering experiment.[30,31] For each spectrum, three vibration modes can be obtained, labeled as $TO_1$ (~60 meV), $TO_2$ (~103 meV), $TO_3$ (~130-150 meV), where $TO_1$ mode is associated with Si-O-Si rocking motions, with bridging oxygen moving perpendicular to the Si-O-Si plane; $TO_2$ mode is assigned to transverse-optical stretching motions, with oxygen atoms moving along a line bisecting the Si-O-Si bond; $TO_{3\text{-surface}}$ mode is associated with Si-O-Si antisymmetric stretching motion, with oxygen atoms moving along a line parallel to the Si-Si axis; as shown in Fig 1c;[31] $TO_{3\text{-bulk}}$ mode is associated with the network disorder.[32] The spatially resolved spectra exhibit these modes have different energy and intensity at different regions, indicating the vibration properties are sensitive to the surface and interface, with the details discussed below.

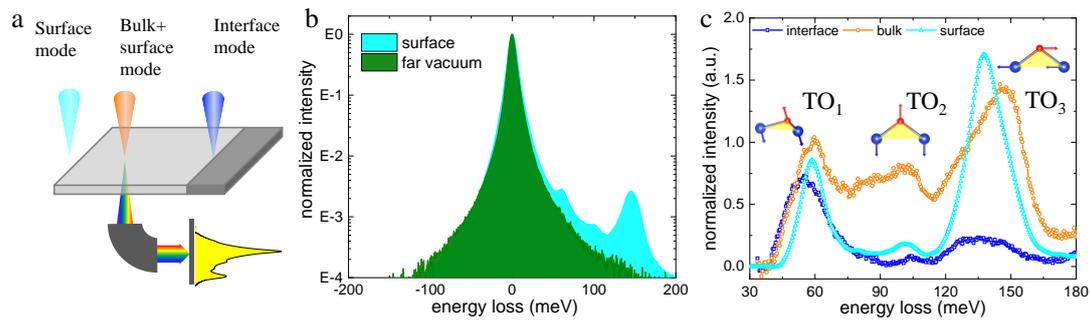

**Fig 1.** Electron energy loss spectroscopy measurement of $SiO_2$ vibration modes in a monochromatic aberration corrected electron microscope. (a) The spectra recorded from different regions represents different vibration modes, surface mode (cyan), bulk + surface mode (orange), interface mode (blue). (b) Normalized spectra acquired with the electron beam located near surface in vacuum side (cyan) and vacuum (green) far away from specimen with energy resolution ~8 meV (characterized by the full width at half maximum). (c) Typical spectra corresponding to (b) after background subtraction. The insets show the schematic representation of the vibrational motions: $TO_1$ mode, rocking motion perpendicular to Si-O-Si plane; $TO_2$ mode, symmetric stretching motion along the bisector of the Si-O-Si bridging angle; $TO_3$ mode, antisymmetric stretching motion parallel to the Si–Si line between the two bridged cations. The orange region shows the Si-O-Si plane. The blue and red arrows indicate the atom motion directions.

We first investigate the surface effect on the vibrational spectra. The translational symmetry is broken at the surface (interface) and a surface (interface) phonon may occur. Generally, the resonance condition is $\varepsilon_1 + \varepsilon_2 = 0$.[33] Fig 2a shows a high angle annular dark field (HAADF) image of an a-$SiO_2$ film with thickness of ~300 nm on Si substrate. In Fig 2b, the selected line profiles of spectra recorded from the vacuum to a-$SiO_2$ with 6.3 nm every step displays the peak of $TO_3$ mode in an a-$SiO_2$ region becomes broad and shifts to higher energy, suggesting the possible co-existence of two vibration modes with different energy in the film. In fact, this is because when placing probe beam inside the a-$SiO_2$ the surface mode (originates from upper and lower surfaces) and the bulk mode are mixed together due to the small difference in energy, broadening the $TO_3$ peak. Fig 2c and Fig 2d are the corresponding EELS mapping for experiments and simulation, respectively, in which the surface position is set to zero, the negative distance denotes that the electron beam is in the vacuum and positive distance denotes that the electron beam is positioned inside the a-$SiO_2$ film. The experiment and simulation both show that the $TO_1$ mode and $TO_3$ mode can spread over ~600 nm in the vacuum, while the $TO_2$ mode only concentrates near the surface. To extract the details of $TO_3$ mode, by using Gaussian functions (see Supporting Fig S2 for details) we decompose the signal in 120 meV~170 meV into two peaks, i.e., $TO_{3\text{-surface}}$ and $TO_{3\text{-bulk}}$, which represent the surface and bulk modes, respectively (the surface mode is 13 meV lower than corresponding bulk mode). For the $TO_1$ mode, the difference between bulk and surface modes is subtle (about 3 meV larger in $SiO_2$ than that in vacuum) which is also consistent with analysis of energy loss functions, while for the TO$_2$ mode, the significant broadening makes it hard to identify the peak positions in the SiO$_2$.

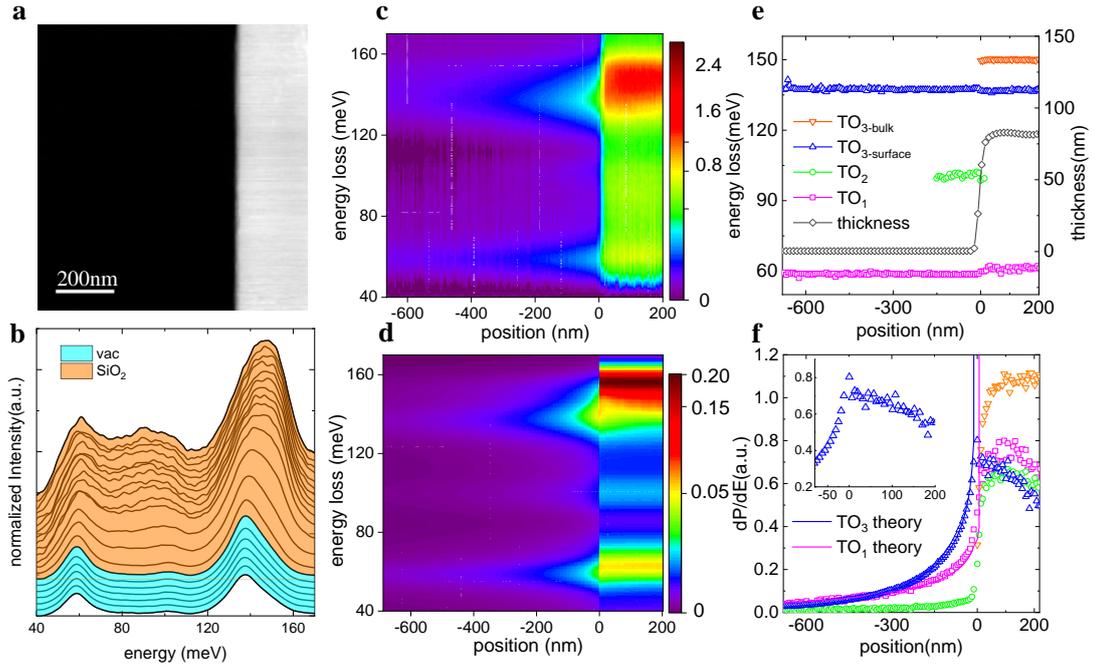

**Fig 2.** Vibrational energy and intensity of $SiO_2$ surface. (a) HAADF image of vacuum/a-$SiO_2$/Si specimen viewed in cross-section. (b) Selected line profiles of the $SiO_2$ vibrational signal with 6.3nm every stack. Orange region, beam located inside the $SiO_2$; cyan region, beam located in vacuum. (c), (d) Two dimensional plots of normalized intensity for experiment and simulation, respectively. The surface position is set to be zero, and negative distance denotes that the electron beam is in vacuum while the positive distance denotes that the beam is positioned within the $SiO_2$. Color bar represents the intensity. (e) Vibration energy of different modes with different color, and z-thickness (black) of $SiO_2$ slab are plotted as a function of the probe position. (f) Vibration intensity as a function of the probe position. Solid line represents theoretical calculation.

The energy of four vibration modes (TO$_1$, TO$_2$, TO$_{3\text{-surface}}$, TO$_{3\text{-bulk}}$,) as well as the film thickness (see Supporting Fig S3 for calculation details) are plotted in Fig 2e. Note that the energy of bulk modes doesn't change. When the electron beam is positioned at surface, the vibration energy of TO$_1$, TO$_2$, TO$_{3\text{-surface}}$ TO$_{3\text{-bulk}}$ modes are 59 meV, 100 meV, 137 meV, 150 meV respectively. The energies of TO$_1$, TO$_2$, TO$_3$ modes are in agreement with the calculation results: 60.5 meV for TO$_1$, 100.4 meV for TO$_2$ and 139 meV for TO$_{3\text{-surface}}$ (Fig 2d). The energy of TO$_{3\text{-bulk}}$ is in agreement with precious reports.[32,34] It should be noted that the TO$_2$ mode 'disappears' when the probe is ~150 nm apart from the SiO$_2$ surface because the intensity is too weak to extract from the tail of the TO$_1$ peak.

Fig 2f shows the normalized intensity as a function of distance to the surface. The intensity of bulk mode TO$_{3\text{-bulk}}$ reaches 80% of its maximum value at 20 nm from the surface. The intensity of the surface mode TO$_{3\text{-surface}}$ has the maximum value at the edge of surface and gradually decrease away from the edge (inset in Fig 2f). This can be understood by the geometry of thin foil structure, since the total intensity of the surface mode is contributed by the up surface, bottom surface, and edge surface. At the edge, all of them make significant contributions while far away from the edge only the up and bottom surface make contributions. The intensity distribution of the TO$_{3\text{-surface}}$ in the vacuum can be described by the relativity-modified dielectric response theory,[35] based on which the inelastic-scattering probability per unit energy loss E per unit length t for aloof-beam spectroscopy is

$$\frac{dP}{dEdt} = \frac{4}{\pi} \frac{\alpha}{\hbar c \beta^2} \text{Im}\left(\frac{-1}{\varepsilon_{SiO_2}(E)+1}\right) K_0\left(\frac{2bE}{\gamma \hbar c \beta}\right) \quad (1)$$

where $t$ is the specimen thickness, α = 1/137 is the fine-structure constant, $\hbar$ is the reduced Planck constant, β = v/c is the ratio of beam velocity v to the speed of light c, $\gamma = (1-\beta^2)^{-1/2}$ is the Lorentz factor, b is the impact parameter (the distance between the beam and the sample surface). The function $K_0$ is the zero-order modified Bessel function of second kind, which represents the dependence of the vibration signal on the impact parameter b. The vibration intensity of surface mode $TO_{1-surface}$ and $TO_{3-surface}$ is very well described by equation (1). The Blue and Red solid line is the theoretical variation derived from Eq(1). The decay behavior can also roughly be fitted with the exponential function: $I = I_0 exp(-b/b_0)$ (ref.[11,36]). With this approximation, the decay distance is fitted to be ~400 nm for $TO_{1-surface}$, ~350 nm for $TO_2$, and ~250 nm for $TO_{3-surface}$ (detailed data see Supporting Table 1). In bulk a-$SiO_2$, the intensity of 'bulk' signal remains constant due to the nearly uniform thickness of the specimen.

When the thickness of specimen is in the order of v/ω (where ω is the frequency of vibration and v is electron beam velocity) or less, the coupling of upper and lower surfaces would give rise to symmetric and asymmetric guided modes.[27] Fig 3a, 3b shows the HAADF image, selected line profiles of $SiO_2$ spectra (4.7 nm every step) with non-uniform thickness, respectively. Fig 3c, 3f show the two-dimensional plots of normalized intensity for experiments and simulations. As the beam is positioned from the edge to the inner, the energy of the $TO_{3-surface}$ mode (blue line) shows significant blue-shift while the $TO_{1-surface}$ shows smaller blue-shift, as shown quantitatively in Fig 3d and Fig 3e. It should be noted that the vibration energy of bulk mode $TO_{3-bulk}$ doesnot change as thickness increases.

We attribute the blue-shift of surface mode to the reduced strength of coupling when z-thickness increases. This energy shift behavior can be interpreted by the thickness dependent energy-loss functions plotted in Fig 3g based on Kröger formula.[27] There are two important features. Firstly, the vibration energy of $TO_{1-surface}$ and $TO_{3-surface}$ modes increases as the specimen thickness increases, while the energy of $TO_2$ mode almost remains unchanged. Consistently, the quantitative analysis in Fig 3d and Fig 3e confirms that different vibration modes have different thickness dependent behavior. Specifically, the energy of the $TO_{3-surface}$ mode at the edge is 133 meV and increases to 136 meV distanced ~200 nm from the edge inside the film. In Fig 3h, three data sets from different specimens and calculation based on Kröger formula further confirm that the energy of the $TO_{3-surface}$ mode is very sensitive to $SiO_2$ thickness.

Generally, the energy increases as specimen thickness increases as we expect from the theory calculation. This can be understood by the strong coupling between the upper and the lower surfaces, i.e., the coupling strength of the upper surface and the lower surface becomes weaker as z-thickness increases. The difference between the experiments and theory may be due to the non-uniform specimen instead of ideal wedge-shaped films.

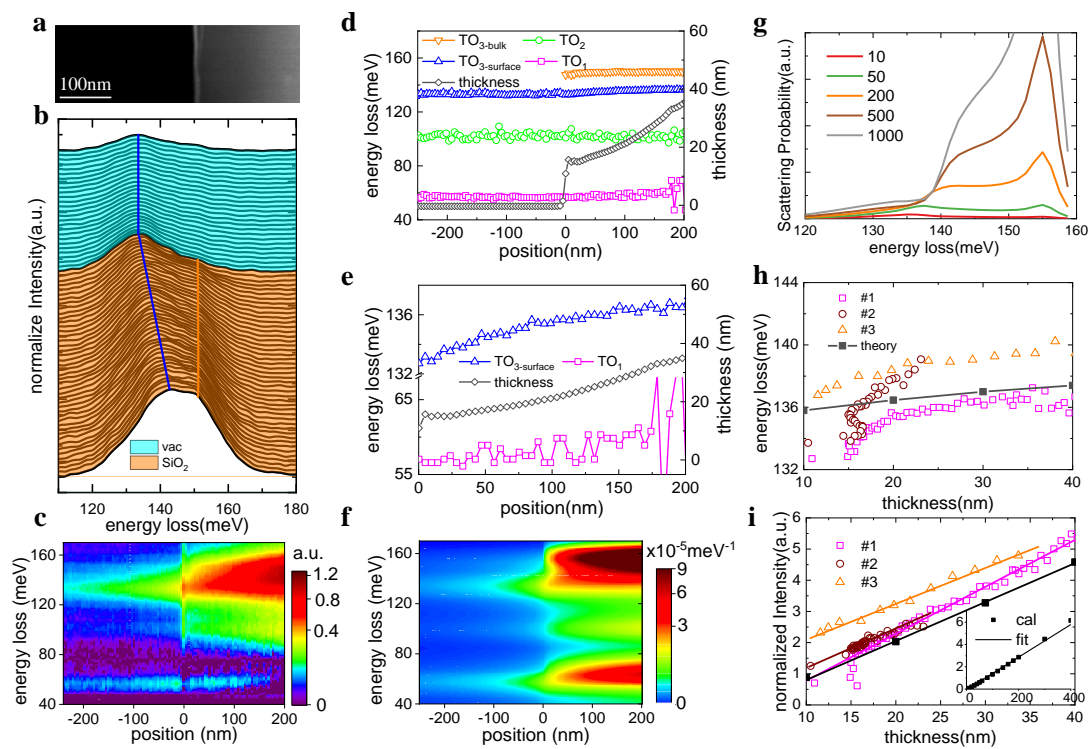

**Fig 3.** Effect of film thickness on the vibrational energy and intensity. (a) HAADF image of vacuum/ a-SiO$_2$ viewed in cross-section. (b) Selected line profile of the SiO$_2$ vibrational signal with 4.7 nm every stack. Orange region, beam located in SiO$_2$; cyan region, beam located in vacuum. Two dimensional plots of normalized intensity for (c) experiments and (f) simulations, respectively. The surface is set to be zero, and negative (positive) distance denotes that the probing beam is in vacuum (SiO$_2$). Color bar represents the normalized intensity. (d) Energy of vibration modes (colored) and z-thickness (black) of SiO$_2$ slab as a function of the probe position. (e) Enlarged view of TO$_{3\text{-surface}}$ mode and TO$_1$ mode showing the peak shift depends on the film thickness. (g) Energy loss function for different thickness. (h) Vibration energy of TO$_3$ mode as a function of z-thickness of SiO$_2$ slab. Colored data are recorded from different samples and the black points are calculated from Kroger formula. (i) The intensity of TO$_3$ mode is plotted as a function of z-thickness of SiO$_2$ slab. Colored data are recorded different samples and black point is calculation from Kroger formula.

The vibration intensity also depends on the thickness of film. Compared to the $TO_{3\text{-surface}}$ mode, the probability of $TO_{3\text{-bulk}}$ mode decreases more quickly as the z-thickness decreases. The normalized intensity of $TO_{3\text{-bulk}}$ mode is plotted as a function of $SiO_2$ slab thickness for the three data sets, which are all proportional to the slab thickness as we expect from the theoretical calculation (shown in inset).

In the following we discuss the interface effect on the lattice vibration. Fig 4a, 4b, shows the atomically resolved HAADF image of $SiO_2$/Si interface, selected line profiles of $SiO_2$ spectra (4.7 nm every step) with non-uniform thickness. Fig 4c and Fig 4e show the two-dimensional plots of normalized intensity for experiments and simulation, respectively. Note that when the beam is located in the $SiO_2$ side, the acquired signal is mixed with the signals of $SiO_2$ (both surface mode and bulk mode) and $SiO_2$/Si interface mode. Fig 4d shows the energy loss function simulated near the $SiO_2$/Si interface, where two vibration modes (interface and bulk) were predicted to obtain. In principle, the simulation shows that the pure signal of $SiO_2$/Si interface vibration mode can be obtained by placing the electron beam in the Si substrate for long-range Coulomb interaction, because the retardation effects mask any inherent vibrational signal of Si substrate.[37] Unfortunately, the signal in Si side is too weak to record even after very long acquisition.

The Surface vibration for 80 nm thick is 137 meV, similar to interface vibration 136 meV. So, we decompose the signal between 120 meV and 160 meV into two Gaussian peaks as the $TO_{3\text{-bulk}}$ (from the bulk vibration with higher energy) and $TO_{3\text{-surface+interface}}$ (from the interface vibration of $SiO_2$/Si, upper and lower surfaces of $SiO_2$ with lower energy). Their respective energy, intensity and intensity ratio of $TO_{3\text{-surface+interface}}$ to $TO_{3\text{-bulk}}$ ratio are shown in Fig 4f, Fig 4g and Fig 4h, respectively. The energy of interface mode is 136 meV just as we expect from the energy loss function shown in Fig 4d. No significant energy shift of both modes is observed in Fig 4f. The intensity of $TO_{3\text{-bulk}}$ mode reaches 70% of its maximum at 3 nm from the interface and the intensity of $TO_{3\text{-surface+interface}}$ mode drops steeply to zero within ~10 nm from the interface. The black solid line is the expected retarded signal intensity of $TO_{3\text{-bulk}}$, according to the reference,[38] which well describes the experiment data in the $SiO_2$ side.

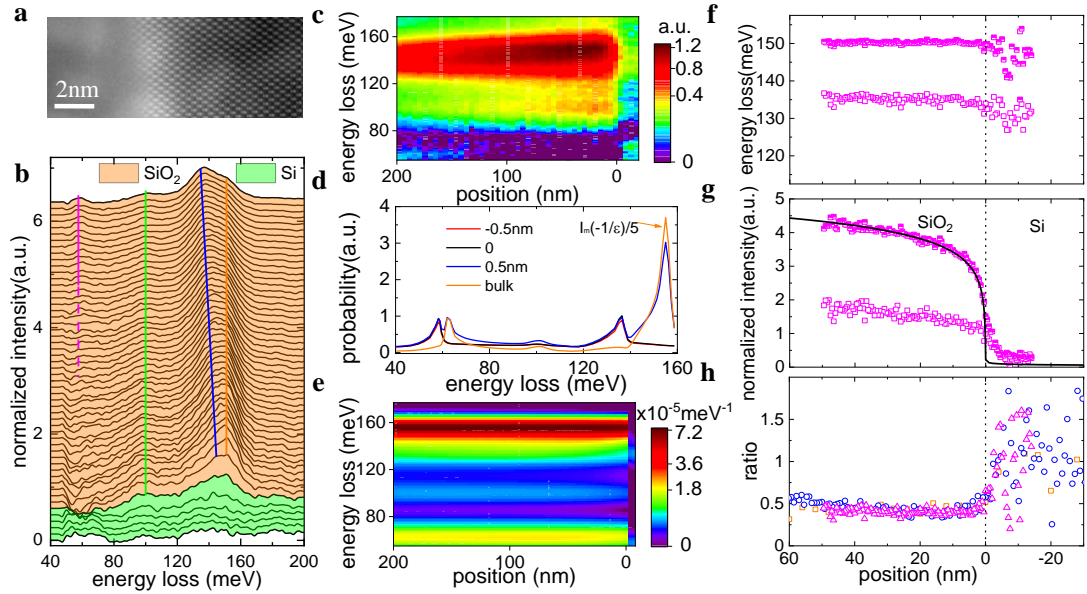

**Fig 4.** Lattice vibration near the interface of $SiO_2$/Si. (a) An atomically resolved HAADF image of amorphous $SiO_2$/Si viewed in cross-section. (b) Line profile of the $SiO_2$ vibrational signal with 0.78nm every stack. Orange region, beam located in a-$SiO_2$ side; green region, beam located in Si side. (c), (e) Two dimensional plots of normalized intensity for experiments and simulations, respectively. The interface is set to be zero, and positive distance denotes that the beam is in the $SiO_2$ while negative distance denotes that the probing beam is positioned in Si. Color bar represents the normalized intensity. (d) Energy loss function based on dielectric theory. (f), (g) The energy and intensity of $TO_{3\text{-bulk}}$ mode and $TO_{3\text{-interface+surface}}$ mode as a function of the probe position; the open and the half up filled squares represent $TO_{3\text{-bulk}}$ mode and $TO_{3\text{-interface+surface}}$ mode, respectively. (h) The intensity of $TO_{3\text{-interface+surface}}$ to $TO_{3\text{-bulk}}$ ratio. The colored data points represent three sets of data recorded from different samples.

The weak vibrational signal in the Si side may be due to the surface oxide layers in Si substrate[37] or the diffused electron probe. From the intensity ratio profile in Fig.4h, the intensity ratio of $TO_{3\text{-surface+interface}}$ to $TO_{3\text{-bulk}}$ in the $SiO_2$ side is almost a constant ~0.5 rather than a bulge at the interface, indicating the contribution from the upper and lower surfaces to total intensity of $TO_{3\text{-surface+interface}}$ is likely dominated due to thin foil geometry. However, in the Si side, the value of ratio is significantly larger than those in the SiO$_2$ side evidenced from three sets of data in Fig.4h. It can be understood by the fact that in the Si side the intensity of both of bulk and surface modes is neglectable and only the interface mode is detected based on the simulation in Fig.4d. Thus, the residual signal in the Si side near the interface mainly represents the interface mode.

**Conclusion**

In summary, by using the monochromatic STEM-EELS in aberration corrected electron microscope, we investigate vibration behavior of surface and interface of SiO$_2$ film on Si substrate. With the improved energy resolution of sub-10 meV in the present study, three optical vibration modes of ~60 meV, ~103 meV and 150 meV are recorded simultaneously. We find that all the surface modes have lower energy compared to the corresponding bulk modes, which is in good agreement with the local dielectric theory. Particularly, the energy of antisymmetric stretching mode is ~13 meV lower for the 60 nm-thick SiO$_2$ slab. The maximum intensity of surface mode occurs at the edge of surface and gradually decrease in the two sides. In the vacuum side, the intensity of surface vibration as a function of impact parameter accurately obey the zero-order modified Bessel function of the second kind[35] with typical decay distances ~400 nm for TO$_{1\text{-surface}}$, ~350 nm for TO$_2$, and ~250 nm for TO$_{3\text{-surface}}$. Inside the SiO$_2$ film, the bulk modes have constant vibration energy while the energy of surface modes strongly depends on the thickness. Approaching the interface of SiO$_2$/Si, a weaker interface mode response at ~136 meV is obtained. All the vibrational signal drops sharply within a few nanometers from the interface in SiO$_2$ side.

Since the properties of the surface and interface of SiO$_2$ films can significantly influence the performance and stability of device,[22,23] the miniaturization of MOSFET's and bipolar transistors in the integrated circuits necessitates the investigation of these boundary conditions. The highly sensitive of vibration properties to the surface, thickness and interface of SiO$_2$/Si is expected to influence the electrical and thermal conductivities in a very complicated manner in MOSFET's and bipolar transistors. Therefore, our experimental measurements of local vibration properties SiO$_2$/Si give useful clues for further understanding and controlling the properties of SiO$_2$/Si heterostructure via surface and interface treatments particularly for those devices with size below 14 nm.

**Acknowledgements**

The work was supported by the National Key R&D Program of China (grant numbers 2016YFA0300804 and 2016YFA0300903); the National Natural Science Foundation of China (Grant Nos. 51502007, 51672007), National Equipment Program of China (ZDYZ2015-1). We gratefully acknowledge the support from the National Program for Thousand Young Talents of China and "2011 Program" Peking-Tsinghua-IOP Collaborative Innovation Center of Quantum Matter. The authors acknowledge Electron Microscopy Laboratory in Peking University for the use of Cs corrected electron microscope.

# Supplemental Material:
# Probing lattice vibration at surface and interface of SiO$_2$/Si with nanometer resolution

Yuehui Li[1], Mei Wu[1], Ruishi Qi[1], Ning Li[1,2], Yuanwei Sun[1], Chenglong Shi[3], Xuetao Zhu[4], Jiandong Guo[4], Dapeng Yu[2,5,6], Peng Gao[1,2,6*]

[1]International Center for Quantum Materials, School of Physics, Peking University, Beijing, 100871, China

[2]Electron Microscopy Laboratory, School of Physics, Peking University, Beijing, 100871, China

[3]Nion Co. 11511 NE 118Th Street, Kirkland, Washington State, 98034, USA

[4]Beijing National Laboratory for Condensed Matter Physics and Institute of Physics, Chinese Academy of Sciences, Beijing 100190, China

[5]Shenzhen Key Laboratory of Quantum Science and Engineering, Shenzhen 518055, China

[6]Collaborative Innovation Centre of Quantum Matter, Beijing 100871, China.

Y.H.L. and M.W. contributed equally to this work.

E-mails: p-gao@pku.edu.cn

**Data acquisition.** In our experiments, we used a Nion UltraSTEM200 microscope with both monochromator and the aberration corrector operating at 60kV. The beam convergence semi-angle was 15mrad and the collection semi-angle was 24.4 mrad with a 1 mm spectrometer entrance aperture. The typical energy resolution (half width of the full zero loss peak, ZLP) was 8meV, the probe beam current was ~5-10 pA and the dispersion of per channel was 0.47 meV. The typical dwell time was 100-200ms to obtain good signal-noise ration spectra. The specimen studied in this work was 300 nm amorphous $SiO_2$ (a-$SiO_2$) on Si substrate. The cross-section sample was prepared by traditional mechanical polishing. We acquired single spectrum from different regions, line scan spectra, and map spectra to resolve the spatially dependent vibration properties.

**Calibration and background subtraction.** To correct the energy shift, we shifted the maximum point of ZLP to zero for unsaturated ZLP and shifted the centroid of two energies whose intensity was 5% lower than its peak value. To obtain more precise information from the spectra, we fitted the background with power law $I(\Delta E) = A_0 \cdot \Delta E^{-r}$. Note that generally, third order exponential provides a more accurate fit, but due to the high energy resolution (~8.3 meV), there is little benefit using the exponential.[1]

**Theoretical analysis.** To quantitatively interpret the experiment results, calculation of the vibration response was performed. There are many theories and models based on classical dielectric response or quantum mechanical exploration to explain nanometer confined vibrations.[2–5] Here we do calculation mainly based on a semiclassical relativistic local dielectric model developed by E. KROGER,[3] which considers the coupling of the upper and lower surface and gives rise to symmetric and asymmetric guided slab modes.

There are two reasons to choose Gaussian function to fit the peak rather than Lorentz function: (1), $SiO_2$ sample is amorphous; a dielectric function based on a modified Gaussian profile was used to reproduce the infrared reflectivity spectra of silicate glass;[6] (2), the probing function (ZLP) is approximated as Gaussian distribution. It should be noted that to get a better decomposition, a new background function was subtracted to make two sides of the TO$_{3\text{-bulk}}$ + TO$_{3\text{-surface}}$ signal approach zero, where two fitting region (105meV~115meV, 175meV~195meV) was chosen.

We calculate film thickness through plasmon peak, using formula:

$$t = \lambda \ln \frac{I_{tot}}{I_{ZLP}}$$

where, $I_{tot}$ is the sum of all counts in a spectrum containing main plasmon signal, $I_{ZLP}$ is the sum of zero loss peak, $\lambda$ is the electron mean free path, whose detailed expression can be obtained from reference.[7]

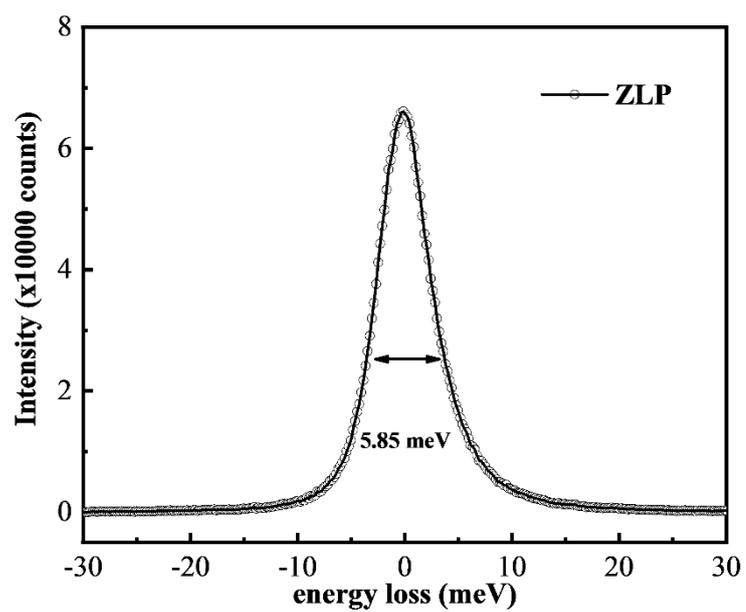

**Fig S1.** The energy resolution is characterized by the FWHM of zero loss peak. The best energy resolution is ~ 6meV.

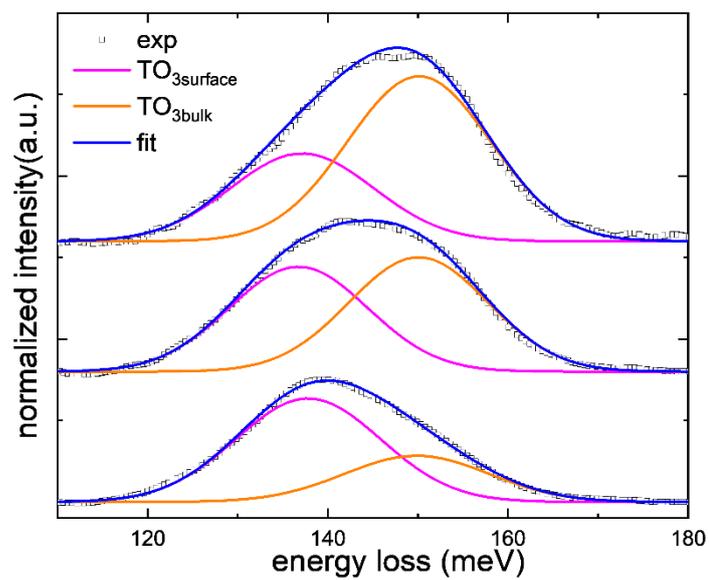

**Fig S2.** Gaussian function to fit the peak. The black open square: experiment data; pink line: the fitted $TO_{3\text{-surface}}$ mode component; orange line: the fitted $TO_{3\text{-bulk}}$ mode component; blue line: sum of $TO_{3\text{-surface}}$ mode component and $TO_{3\text{-bulk}}$ mode component.

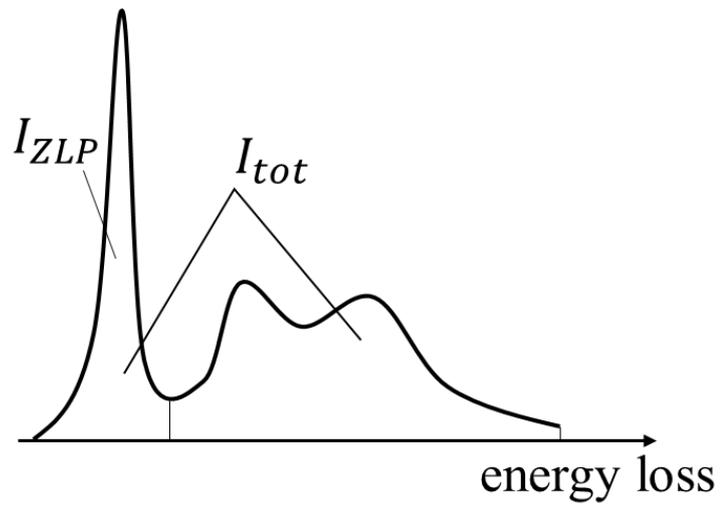

**Fig S3.** Sketch diagram of calculating thickness by log-ratio method.[7]

|                 | #1  | #2  | #3  | #4  |
|-----------------|-----|-----|-----|-----|
| TO$_1$          | 415 | 380 | 394 |     |
| TO$_2$          | 357 | 338 | 333 | 353 |
| TO$_3$          | 259 | 276 | 242 | 251 |

**Table 1.** decay constant b$_0$ (nm) for different vibration modes. Different sets of data were fitted with exponential function $I = I_0 exp(-b/b_0)$.